\documentstyle[twocolumn,prl,aps]{revtex}
\input epsf
\begin{document} 
\twocolumn
\title{
Peak Effect and the Transition from Elastic to Plastic Depinning 
}
\author{
Min-Chul~Cha$^1$ and H.A.~Fertig$^2$
}
\address{
$^1$Department of Physics, Hanyang University, Ansan, Kyunggi-do, 425-791, Korea 
\\
$^2$Department of Physics and Astronomy and Center for
Computational Sciences, University of Kentucky,
Lexington, KY 40506-0055
}
\address{\mbox{ }}
\address{\parbox{14.5cm}{\rm \mbox{ }\mbox{ }
We demonstrate for the first time an observation of the peak effect
in simulations of magnetic vortices in a superconductor.  
The shear modulus $c_{66}$
of the vortex lattice is tuned by adding a fictitious {attractive}
short range potential to the usual long-range repulsion between
vortices.  The peak effect is found to be most pronounced in
low densities of pinning centers, and is always associated with
a transition from elastic to plastic depinning.  The simulations suggest
in some situations
that over a range of values of $c_{66}$ the production of lattice
defects by a driving force {\it enhances} the pinning of the lattice.
}}
\address{\mbox{ }}
\address{\mbox{ }}

\maketitle

One of the most important aspects of superconducting materials
is their behavior in magnetic fields.  In Type II superconductors,
magnetic fields induce the formation of quantized vortices through
which magnetic flux may penetrate the system.   The application of a current
in the presence of vortices generates an effective force 
that causes them to flow through the superconductor,
thereby dissipating energy and spoiling perfect conductivity.
Much work, both theoretical and experimental, has been devoted
over the years to finding and understanding mechanisms that
pin the vortices, so that the superconductor will remain dissipationless
even in the presence of the field\cite{blatter94}.  In the 
presence of pinning centers, the superconductor displays a
critical current $j_c$ above which dissipation sets in; physically
this current is proportional to a critical force $F_p$
at which the vortices become depinned and can move through the system.

The competition between intervortex interactions and
pinning by disorder results in a surprising 
and long studied phenomenon
known as the
``peak effect''.  As the critical field or critical temperature
at which a sample loses its superconducting properties 
is approached, in many
situations one observes an {\it enhancement} of the 
superconductivity just before it is completely
suppressed. Thus, $j_c$ exhibits a
peak as a function of field or temperature, just before it
vanishes.  The earliest understanding of this phenomenon,
the collective pinning theory, involves
the softening of the elastic moduli of the vortex lattice
as the superconducting order is suppressed\cite{pippard69,larkin79},
so that the vortices may settle more deeply into the pinning
potential and thus become more difficult to dislodge.
One difficulty with collective pinning is 
that tearing of the lattice is ignored 
in estimating $j_c$: {\it i.e.}, the lattice depins elastically,
not plastically.  In recent years this assumption
has been increasingly questioned in the
peak effect regime, particularly for high-purity superconductors
(e.g., NbSe$_2$\cite{yaron95,bhatt93,brill94}) and 
systems with strong pinning centers\cite{matsuda96,harada96,twin}.
Although there is accumulating experimental evidence 
of plastic motion in the peak effect regime,
its precise effect on the size of $j_c$ is not
known, and there is disagreement as to whether $j_c$
is enhanced\cite{bhatt93,larkin95} or 
suppressed\cite{twin,pastoriza95} by the onset
of plastic motion.

In principle much of this debate could be settled by
direct imaging of the vortices near the depinning
critical current.  However, in the peak effect regime
such experiments are exceedingly difficult because
the order parameter is suppressed near the critical temperature
or critical magnetic field.  Numerical simulations thus offer
a unique window through which one may view the qualitative
behavior of the 
vortices\cite{nori96,ryu96,koshelev94,moon96,faleski96}.
In this work, we present
results demonstrating for the first time
(albeit, in a two-dimensional geometry) the peak effect
in a simulated vortex system, show conclusively
that it is associated with a crossover from
elastic to plastic motion, and find that under
different circumstances lattice tearing may
enhance or suppress $j_c$.

In recent years, the peak effect has been 
associated with the proximity of the vortex lattice
to a melting transition\cite{larkin95,ghosh96,nelson88}.
Direct simulations of depinning in
this situation pose enormous practical problems
because near a critical point one inevitably has
large thermal fluctuations.  To circumvent
this problem, we take note that in nearly every theoretical
approach to the peak effect, it is not actually melting
itself but rather the softening of $c_{66}$ and other
elastic moduli near the melting transition that is responsible
for the effect.  We thus consider a system of vortices in
which the interaction may be varied so that the 
elastic properties of the system may be tuned directly,
without the introduction of critical fluctuations.
To simulate a large number of vortices ($N_V =1600$) we
confine ourselves to two-dimensional systems, so that
our simulations are most directly applicable to very
thin films
or superconductors consisting of effectively decoupled layers.
In contrast to previous simulations, we focus on pinning centers
that are dilute compared to the vortex density, which is most
appropriate for systems with strong pinning 
centers\cite{matsuda96,harada96,baert95,twin}.

The precise form of the intervortex interaction we use is
\vbox{
\begin{eqnarray}
H_{int}=-{1 \over 2}  \sum_{\vec R \not= \vec R^\prime}
\{e^2 \ln |\vec R - \vec R^\prime|
+ A_v e^{-|\vec R - \vec R^\prime|^2/\xi_v^2}
\} 
\eqnum{1}
\end{eqnarray}
}
where $\vec R$ are the position vectors of vortices, $e^2$ is the strength of
the logarithmic interaction, and $A_v$ and $\xi_v$ are the strength and
the range of a short range attractive interaction.
For large enough values of $A_v$ the interaction
may in principle be attractive over a range of vortex
separations; however, in all the simulations we report here,
$A_v$ is small enough that the net interaction is repulsive
at all distances.
A uniform background is assumed to cancel out the diverging energy due to the
logarithmic interaction.
Because of the long-range potential, the bulk
modulus is formally divergent, while
the shear modulus $c_{66}$ may be shown
to have the form\cite{Alastuey81}
\begin{eqnarray}
c_{66}= n_v \{{e^2 \over 8}
- {A_v \over 2} \sum_{\vec R}
[{1 \over 2} ({R \over \xi_v})^4 - ({R \over \xi_v})^2]
e^{-R^2/\xi_v^2} \}
\eqnum{2}
\end{eqnarray}
where $n_v =2/(\sqrt{3}a_0^2)$ is the density of vortices.
We take $\xi_v =0.5 a_0$ so that the shear modulus has values
$c_{66}= (n_v / 8) (e^2 - 1.768 A_v)$, and tune $c_{66}$
by varying $A_v$.
The vortices interact with the pinning centers through the potential:
\begin{eqnarray}
H_{pin}=-A_p \sum_{\vec R, \vec r}
e^{-|\vec R - \vec r|^2/\xi_p^2}
\eqnum{3}
\end{eqnarray}
where $A_p$ is the strength of the pinning centers and
$\vec r$ are the positions of the pinning centers\cite{com1}.

Systems with $N_V=900$ and $N_V=1600$ vortices and different numbers of
pinning centers located randomly 
are studied by a simulated annealing molecular dynamics (MD) method.
Periodic boundary conditions are imposed and
an Ewald sum technique\cite{Alastuey81}
is used to compute the forces and energies.
To equilibrate the system, the
temperature is lowered from above the melting temperature
to $kT= 0.001 e^2$
in about 30 consecutive steps
through typically $10^5$ MD steps.
The depinning force is then measured at very low temperature,
using a quasistatic technique\cite{berlinsky,cha94}
as follows.  The center of mass of the system is shifted
in steps of 0.01$a_0$ by imposing a driving force. 
At each step, 
the driving force is allowed to fluctuate while the
center of mass is fixed, and 1000 MD steps are allowed
to pass to equilibrate the shifted system.  The average
driving force required to hold the center of mass at this 
position is then measured over 200 MD steps.  The average driving
force increases approximately linearly with center of mass
shift until the system finds a new minimum energy configuration,
at which time the required driving force drops sharply.  The
depinning force is then defined as the peak driving force observed
in this process.

\begin{figure}
\hbox{ \epsfxsize=3.in
\epsffile{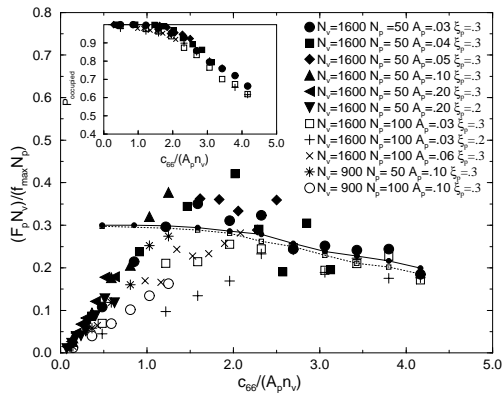}}
\caption{Depinning force $F_p$ as a function of shear modulus
$c_{66}$ for $N_V$ vortices and $N_p$ pinning
centers of strength $A_p$ and range $\xi_p$.  The maximum possible
force ${\rm f_{max}}$ and the number of pinning sites per vortex
$N_p/N_V$ are scaled out so that data should collapse onto
a single curve if the lattice depins elastically.  Solid and
dotted lines illustrate the expected behavior for elastic
depinning.  Inset: Fraction of occupied pinning sites ${\rm P_{occupied}}$
for the groundstate configurations found by simulated annealing.
}
\label{fig:wk_pinning}
\end{figure}

Fig. \ref{fig:wk_pinning} illustrates our results for the depinning threshold
force $F_p$
as a function of the shear modulus 
$c_{66}$ for several different choices of
$\xi_p$, $A_p$, and $N_p$, the number of pinning centers.
In our simulations, we find that the peak effect is
most pronounced when the density of pinning centers
is small compared to the number of vortices, so we focus 
our attention on simulations
with $N_p=50$ and 100.   
For increasing values of $c_{66}$, we expect the number
of pinned vortices to decrease, as in collective pinning.
This general trend is confirmed
in the inset of Fig. 1, which illustrates the fraction of
occupied pinning sites ${\rm P_{occupied}}$, defined as the
fraction of sites for which a vortex may be found within
$\xi_p/\sqrt{2}$ of the pinning site center.  When the
lattice depins elastically (i.e., when tearing and defect
formation may be ignored),
one expects the threshold depinning force to be proportional
to the density of occupied pinning sites and the maximum
pinning force that a site may exert,
${\rm f_{max}}= (A_p / \xi_p )\sqrt{2} e^{-1/2}$.
In the main part of Fig. \ref{fig:wk_pinning}, 
${\rm f_{max}}$ has been scaled out of $F_p$, and
two curves proportional to ${\rm P_{occupied}}$
are plotted for the $N_p=50,~100$ data with the proportionality
constants chosen to match the threshold force
for the largest values of $c_{66}$,
where the depinning is most elastic.
As may be seen, the pinning force matches
the expectations for elastic depinning reasonably well down to
$c_{66}/A_pn_v \approx 2.0$.
For $N_p=100$, $F_p$ decreases monotonically
to zero as $c_{66} \rightarrow 0$, whereas for $N_p=50$ there
is a clear tendency for $F_p$ to {\it overshoot} the elastic
depinning estimate before dropping to zero.
This non-monotonic behavior
of $F_p$ vs. $c_{66}$ is the peak effect, and one of the
surprising results of this study is
that it is more pronounced for lower densities
of pinning centers.  
This latter result is in qualitative agreement with
experiment, for which samples that are more weakly
disordered exhibit stronger peak effects\cite{bhatt93}.

The deviations from the elastic depinning estimate occur because
for small values of $c_{66}$, the lattice easily deforms
and tears, allowing motion without 
forcing out all the vortices trapped in pinning sites. 
Fig. \ref{fig:trajectories} illustrates the trajectories of
the vortices for large, intermediate, and small values
of $c_{66}$.  For the largest values, the vortex lattice
largely retains its order as it depins.  As the maximum of the peak
in $F_p$ is approached, a crossover from elastic to
more plastic motion is observed, in which dynamically
changing channels form where
vortex motion takes place.  The width 
of these channels decrease with
decreasing $c_{66}$.  This behavior is reminiscent
of plastic motion observed in superconductors with ordered
arrays of pinning centers\cite{matsuda96} and in simulations
of disordered Wigner crystals\cite{cha94}.  As the falling
edge of the peak effect is entered, a new qualitative
behavior emerges in which the active channels of motion
of the vortices are no longer dynamic, and the motion
becomes very much like river flow\cite{watson97}.

Scenarios in which the peak effect is associated with
a crossover from elastic to plastic motion have been 
advocated by several groups in the last few 
years\cite{bhatt93,twin,larkin95}.
The present simulations strongly support this viewpoint,
although the precise evolution of the flows with $c_{66}$
differs in some important aspects from what previously
has been supposed.  In particular the 
onset of plastic motion in the peak effect
regime has been thought to be associated
with either a monotonically increasing\cite{bhatt93,larkin95}
or decreasing\cite{twin,pastoriza95} $F_p$ with decreasing
$c_{66}$.  Our simulations demonstrate that in a 
sense both scenarios are true.  Plastic flow,  
when it first sets in with decreasing $c_{66}$, is associated
with an increasing critical current.  This is particularly true
for very low densities of pinning centers, for which $F_p$ is 
{\it enhanced} by tearing\cite{scenario}.  
For low enough values of $c_{66}$, however,
river flow motion sets in and
$F_p$ becomes proportional to $c_{66}$.
The latter behavior is quite sensible once
one recognizes from the simulations that the motion of vortices
for the smallest values of $c_{66}$ correspond to
river flow through channels that do {\it not} change
dynamically. In this case the depinning force comes
about due to interactions of the ``rivers'' (moving vortices)
with the ``river banks'' (stationary vortices), whose
ability to hold the rivers in place  decreases
with decreasing shear modulus.

A useful way of characterizing the depinning force
for the smallest values of $c_{66}$
(the ``static river'' limit)
is to assume that if few of the vortices
trapped in pinning sites are pulled free by the
depinning driving force, then the only relevant
length scale in the system at the depinning transition
is $d \propto ({n_v} / {n_p})^{1/2}$, 
the average distance
between pinning sites.  In particular this implies
a characteristic displacement scale 
for the lattice in the presence of the driving force
$u_{max} \equiv r_c d$,
with $r_c$ a unitless constant,
above which defects are produced so that
the lattice becomes depinned.  The work done by
the driving force must provide the elastic deformation
energy just before defects and depinning set in,
so that $F_pn_v r_c d \sim c_{66} r_c^2 d^2$, or
$F_pn_v \propto c_{66}(n_v/n_p)^{1/2}.$  Fig. 3
illustrates this scaling relation,
and one may clearly see a collapse of the data
onto a single straight line for $c_{66}/A_pn_v < 1.0$.
The collapse of the data indicate that $d$ is indeed
the only relevant length scale in the static river flow regime.
%
We note finally that in our limit
of dilute pinning centers, river flow motion is
possible for an unmelted vortex system ($c_{66} > 0$),
in contrast to what has been speculated for
systems with dense pinning centers\cite{blatter94,bhatt93}. 

In summary we have reported the first simulations of the
peak effect in a vortex lattice.   
We observe a peak in the depinning force
near the smallest values of $c_{66}$, demonstrate 
with particle trajectories that this peak is
associated with a crossover from elastic to plastic
motion, and find that the peak is most pronounced for low pinning center
densities.

This work was supported by NSF Grant No. DMR95-03814
and a Cottrell Scholar Award of Research Corporation (HAF),
and by Korea Science and Engineering Foundation 961-0202-008-2 and 
BSRI-97-2448 (MCC).
\smallskip

\noindent {\it Note added:} After the submission of this work,
a publication\cite{gammel} appeared reporting experiments on
Nb using neutron scattering to measure correlation
lengths of a vortex lattice.  It was found that in the 
peak effect regime, the correlation length corresponding
to shear displacements decreases monotonically to a minimum through
the rising edge of the peak effect.  
This observation corroborates our finding
that the softening of $c_{66}$ may be the controlling
parameter in the
peak effect.

\begin{figure}
\caption{Trajectory plots for depinned vortices at different
values of $c_{66}$, illustrating the evolution from elastic
to plastic motion.
Crosses represent the locations of pinning centers.
$A_p=0.03 e^2$, $N_V=1600$, and $N_p=50$, and
(a) $c_{66}/(A_p n_v)=4.17$;
(b) $c_{66}/(A_p n_v)=1.96$;
(c) $c_{66}/(A_p n_v)=0.48$.
}
\label{fig:trajectories}
\end{figure}

\begin{figure}
\hbox{ \epsfxsize=3.in
\epsffile{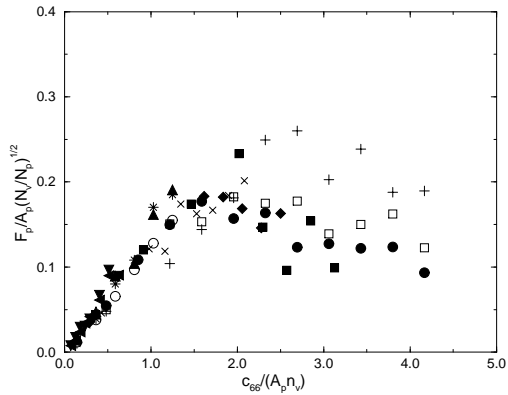}}
\caption{Depinning force $F_p$ times inter-pinning-center spacing
as a function of shear modulus $c_{66}$.
The symbols represent parameters as used in Fig.~1.
The figure clearly shows that in the region where $c_{66}$ is small
$F_p$ is proportional to $c_{66}$, and that the average distance
between pinning centers is the only relevant length scale
for ``static river flow'' depinning (see text).
}
\label{fig:st_pinning}
\end{figure}


\begin{references}
\bibitem{com1} Since the softening of lattice here is due to thermal
fluctuations rather than a mean-field decrease in the order
parameter, it is appropriate to consider a decreasing $c_{66}$
even as the pinning force and long-range components of
the interaction are kept fixed.
\bibitem{scenario} A scenario that can explain
this behavior is that defects developing in the lattice
as the driving force is increased from zero actually allow
a significant fraction of
pinned vortices to settle more deeply into the pinning
centers, thereby increasing the net force needed to drive
enough of them out to establish a flow of vortices.  Work
is currently underway to test this idea.
\end{references}
\end{document}